# The Earth System Grid: Supporting the Next Generation of Climate Modeling Research

DAVID BERNHOLDT, SHISHIR BHARATHI, DAVID BROWN, KASIDIT CHANCHIO,
MEILI CHEN, ANN CHERVENAK, LUCA CINQUINI, BOB DRACH, IAN FOSTER,
PETER FOX, JOSE GARCIA, CARL KESSELMAN, ROB MARKEL, DON MIDDLETON,
VERONIKA NEFEDOVA, LINE POUCHARD, ARIE SHOSHANI, ALEX SIM, GARY STRAND,
AND DEAN WILLIAMS

*Invited Paper*

*Understanding the earth's climate system and how it might be changing is a preeminent scientific challenge. Global climate models are used to simulate past, present, and future climates, and experiments are executed continuously on an array of distributed supercomputers. The resulting data archive, spread over several sites, currently contains upwards of 100 TB of simulation data and is growing rapidly. Looking toward mid-decade and beyond, we must anticipate and prepare for distributed climate research data holdings of many petabytes. The Earth System Grid (ESG) is a collaborative interdisciplinary project aimed at addressing the challenge of enabling management, discovery, access, and analysis of these critically important datasets in a distributed and heterogeneous computational environment.*

*The problem is fundamentally a Grid problem. Building upon the Globus toolkit and a variety of other technologies, ESG is developing an environment that addresses authentication, authorization for data access, large-scale data transport and management, services and abstractions for high-performance remote data access, mechanisms for scalable data replication, cataloging with rich semantic and syntactic information, data discovery, distributed monitoring, and Web-based portals for using the system.*

Manuscript received March 1, 2004; revised June 1, 2004. This work was supported in part by the U.S. Department of Energy under the Scientific Discovery Through Advanced Computation (SciDAC) Program Grant DE-FC02-01ER25453.

D. Bernholdt, K. Chanchio, M. Chen, and L. Pouchard are with the Oak Ridge National Laboratory, Oak Ridge, TN 37831 USA (e-mail: bernholdtde@ornl.gov; chanchiok@ornl.gov; chenml@ornl.gov; pouchardlc@ornl.gov).

S. Bharathi, A. Chervenak, and C. Kesselman are with the USC Information Sciences Institute, Marina del Rey, CA 90292 USA (e-mail: shishir@isi.edu; annc@isi.edu; carl@isi.edu).

D. Brown, L. Cinquini, P. Fox, J. Garcia, R. Markel, D. Middleton, and G. Strand are with the National Center for Atmospheric Research, Boulder, CO 80305 USA (e-mail: dbrown@ucar.edu; luca@ucar.edu; pfox@ucar.edu; jgarcia@ucar.edu; don@ucar.edu; strandwg@ucar.edu).

B. Drach and D. Williams are with the Lawrence Livermore National Laboratory, Livermore, CA 94550 USA (e-mail: drach@llnl.gov; williams13@llnl.gov).

I. Foster and V. Nefedova are with the Argonne National Laboratory, Argonne, IL 60439 USA (e-mail: foster@mcs.anl.gov; nefedova@mcs.anl.gov).

A. Shoshani and A. Sim are with the Lawrence Berkeley National Laboratory, Berkeley, CA 94720 USA (e-mail: shoshani@lbl.gov; asim@lbl.gov).

Digital Object Identifier 10.1109/JPROC.2004.842745

## I. Introduction

Global climate research today faces a critical challenge: how to deal with increasingly complex datasets that are fast becoming too massive for current storage, manipulation, archiving, navigation, and retrieval capabilities. High-resolution and long-duration simulations performed with more advanced earth-system components (e.g., the atmosphere, oceans, land, sea ice, and biosphere) produce petabytes of data. To be useful, this output must be made easily accessible by researchers nationwide, at national laboratories, universities, other research laboratories, and other institutions. Thus, we need to create and deploy new tools that allow data *producers* to publish their data in a secure manner and that allow data *consumers* to access that data flexibly and reliably. In this way, we can increase the scientific productivity of U.S. climate researchers by turning climate datasets into community resources.

The goal of the Earth System Grid (ESG) project [1] is to create a virtual collaborative environment linking distributed centers, users, models, and data. The ESG research and development program was designed to develop and deploy the technologies required to provide scientists with virtual proximity to the distributed data and resources that they need to perform their research. Participants in ESG include the National Center for Atmospheric Research (NCAR), Boulder, CO; Lawrence Livermore National Laboratory (LLNL), Livermore, CA; Oak Ridge National Laboratory (ORNL), Oak Ridge, TN; Argonne National Laboratory (ANL), Argonne, IL; Lawrence Berkeley National Laboratory (LBNL), Berkeley, CA; USC Information Sciences Institute (USC-ISI), Marina del Rey, CA; and, most recently, Los Alamos National Laboratory (LANL), Los Alamos, NM.

Over the past three years, ESG has made considerable progress towards the goal of providing a collaborative environment for Earth System scientists. First, we have developed a suite of metadata technologies including standard schema, automated metadata extraction, and a metadata catalog service. Second, we have developed and deployed security technologies that include Web-based user registration and authentication and community authorization. Data transport technologies developed for ESG include grid-enabled data transport for high-performance access; robust multiple file transport; integration of this transport with mass storage systems; and support for dataset aggregation and subsetting. Finally, we have developed a Web portal that integrates many of these capabilities and provides interactive user access to climate data holdings. We have cataloged close to 100 TB of climate data, all with rich scientific metadata, which provides the beginnings of a digital scientific notebook describing the experiments that were conducted. We have demonstrated a series of increasingly functional versions of this software at events such as the NCAR Community Climate System Model (CCSM) [2] workshop and the Supercomputing (SC) conference [3], and we deployed a system in Spring 2004 that will provide community access to assessment datasets produced by the Intergovernmental Panel on Climate Change (IPCC) [4].

ESG aims to improve greatly the utility of shared community climate model datasets by enhancing the scientist's ability to discover and access datasets of interest as well as enabling increased access to the tools required to analyze and interpret the data. Current global climate models are run on supercomputers within the U.S. and beyond (e.g., the Earth Simulator Center in Japan) and produce terabytes of data that are stored on local archival storage. The analyses of these data contribute to our understanding of our planet, how we are influencing climate change, and how policy makers react and respond to the scientific information that we, as a global community, produce.

Future trends in climate modeling will only increase computational and storage requirements. Scientific advances will require a massive increase in computing capability and a sublinear but still extremely substantial increase in the volume and distribution of climate modeling data. We will see increases in the physical resolution of the models, an elevation of the number of ensemble runs, enhanced "quality" in terms of clouds, aerosols, biogeochemical cycles, and other parameters, and a broadening of overall scope that extends into the upper-atmosphere regions. If we project into the future, it is clear that our global earth system modeling activities are going to produce petabytes of data that are increasingly complex and distributed due to the nature of our computational resources.

The climate modeling community has been creating shared organization- or project-specific data archives for some time. However, while these archives help enable the work of specific well-defined groups, they limit access to the community at large and prohibit analysis and comparison across archives. With the current infrastructure, analysis of these large distributed datasets that are confined to separate storage systems at geographically distributed centers is a daunting task. ESG's goal is to dramatically improve this situation. By leveraging emerging developments in Grid technology, we can break down these artificial barriers, creating a data-access environment that encompasses multiple computational realms spanning organizational boundaries with the goal of delivering seamless access of diverse climate modeling data to a broad user base. This task is challenging from both the technical and cultural standpoints.

In this paper, we describe what the ESG project has accomplished during the last three years and discuss ongoing and future efforts that will increase the utility of the ESG to the climate community over the coming decade.

## II. FUNCTIONAL OBJECTIVES

Next, we present functional objectives and use cases related to climate model dataset management and access. Various climate model simulations have been performed, and the output from these simulations has been stored on archival storage, with each simulation output comprising several thousand files and several hundred gigabytes of data. The resulting data are of great interest to many researchers, policy makers, educators, and others. However, with current technology, not only must the data manager spend considerable time managing the process of data creation, but any user who wishes to access this data must engage in a difficult and tedious process that requires considerable knowledge of all the services and resources necessary: metadata for search and discovery, specifics of the archival system software, system accounts, analysis software for extracting specific subsets, and so on. Because of this complexity, data access tends to be restricted to privileged specialists.

The goal of the ESG system is to simplify both the data management task, by making it easier for data managers to make data available to others, and the data access task, by making climate data as easy to access as Web pages via a Web browser.

### A. Making Data Available to the ESG Community

A first requirement is for tools that allow climate model data managers to make model data available to the ESG community. These tools include the means to create searchable databases of the metadata, provide catalogs of the data that locate a given piece of data on an archival system or online storage, and make catalogs and data accessible via the Web. Prior to the advent of ESG, these capabilities did not exist, so potential users of the model data had to contact the data managers personally and begin the laborious process of retrieving the data they wanted.

It is important that tools for data publishers and curators are robust and easy to use. Data publishers should not have to be familiar with all the implementation details of the system. Thus, the tools must be intuitive, reflect the user's perspective of the data, and be accessible on the user's local workstation.

Although climate simulation datasets are relatively static, they can change over time as errors are corrected, existing runs are extended, and new postrun analyses are added. Thus, it is not sufficient just to be able to publish a dataset once. It must be easy for publishers to update metadata over time. The publishing tools must allow sufficiently privileged users to search and update the metadata database, ensuring that the metadata and physical datasets stay synchronized.

### B. Defining Virtual Datasets

In addition to the "physical" datasets that are generated directly by climate simulations and maintained on archival storage, data producers want to define *virtual datasets*, datasets that are defined by a set of transformation operations on other physical or virtual datasets. These virtual datasets can then be instantiated to yield *physical datasets* by the ESG system, either proactively or in an on-demand fashion following a user request. Physical datasets can be discarded following use or, alternatively, cached to avoid the cost of subsequent regeneration. Virtual datasets are important, as they allow data producers to define abstractions over the physical files that may be more convenient and/or efficient for users to access.

The transformations used to define virtual datasets may include concatenation (of files from one or more datasets), subsetting (along one or more dimensions), and/or computation (e.g., reduction operations). In our work in ESG, we consider only concatenation and subsetting, but our architecture can be extended in the future to support arbitrary transformations, as in the Chimera system [5]. The following is an example of a virtual dataset defined via concatenation and subsetting: "A dataset that contains, from each of two datasets file1.nc and file2.nc, the 'PS' field for 10 time periods." The net effect is to provide the data consumer with a virtual view of a dataset that hides the underlying organization and complexity of thousands of related files.

### C. Providing Simple, Convenient Web-Based Access

A member of the ESG community should be able to browse, search, discover, and access distributed climate model data easily using the ESG Web portal [1]. She can browse the ESG data holdings hierarchically, and a simple text search capability allows her to perform metadata-based searches of data catalogs. Subsequent to browsing and/or search, she can select the data she wants, after possibly narrowing her choices via additional possibilities presented by the search results. She can also select individual fields and/or regions of interest, specifying a selection that will ultimately come from multiple physical files. The Web portal creates an efficient means of finding and accessing published model data; the user states what she wants to access and ESG performs the tasks required to deliver the desired result.

### D. Retrieving Large or Numerous Datasets

If an ESG user wishes to access large amounts of climate model data, particularly data located on an archival system, ESG has a tool called DataMover [6] that efficiently and robustly accomplishes this task. For example, an ESG user may wish to access many simulated years of model data and store that data on his local online storage. DataMover allows the user to move this large volume of data without requiring him to personally monitor all the steps involved. It manages the processes and takes corrective actions as needed to transmit the data from the archival storage system to a location of the user's choosing.

### E. Using a Climate Analysis Application to Access ESG Holdings

An ESG user may choose to utilize a common data analysis package like Climate Data Analysis Tools (CDAT) [7] or NCAR Graphics Command Language (NCL) [8] to access the data published via ESG. For example, a Web portal search will give her the information she needs to pull the data directly into her software package. This implies a requirement for interfaces that are compatible with these applications and distributed data access protocols that provide for remote access. Such a capability allows the ESG user community far greater access to climate model data than was previously available as well as the ability to do extensive and complicated analyses on large sets of data efficiently, since there is no need to retrieve and store all the data locally before beginning analysis.

## III. ESG ARCHITECTURE

Next, we present a description of the overall ESG architecture. Fig. 1 shows the major components involved in the ESG system, which are, from bottom to top, the following.

- *Database, Storage, and Application Servers:* These are the basic resources on which ESG is constructed and include the computational resources used for creation of virtual data, disk caches and mass storage systems used to hold data assets, database servers used to store metadata, and application servers for hosting the ESG portal services.
- *Globus/Grid infrastructure*: This provides remote, authenticated access to shared ESG resources, enabling access to credentials, data, and metadata, replica location services (RLSs), and the submission and management of computational jobs.
- *High-level and ESG-specific services*: These services span ESG resources and provide capabilities such as site to site data movement, distributed and reliable metadata access, and data aggregation and filtering.
- *ESG applications*: The most important of these is the ESG portal, which provides a convenient, browser-based interface to ESG services. Other applications include user level tools for data publication

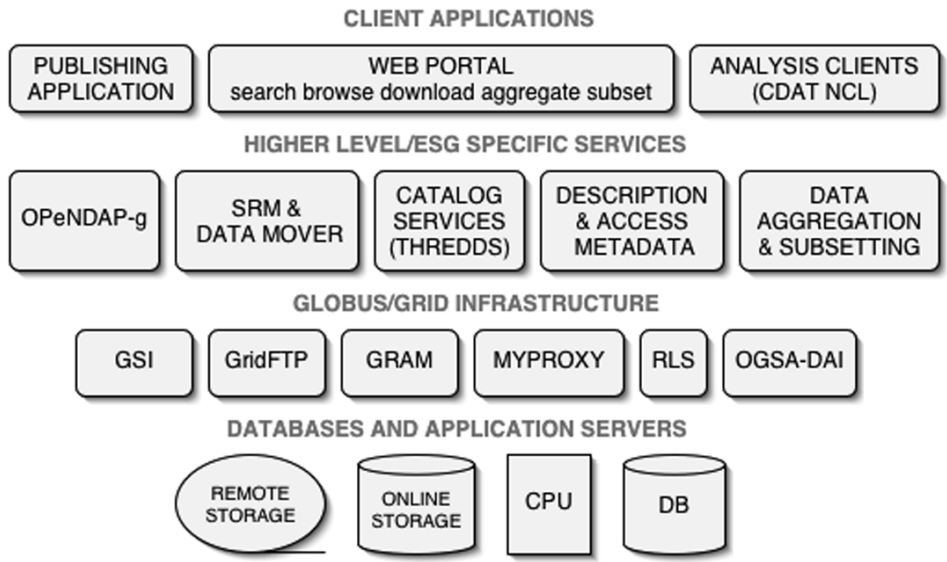

**Fig. 1.** ESG architecture schematic showing major components.

as well as assorted clients for data analysis and visualization.

## IV. Accomplishments: Bringing Climate Datasets to the User Community

During our first three years of ESG development, we have reached milestone achievements in the following areas:

- speed of access and transport of data at user request, via the integration of advanced GridFTP data transfer technologies into the widely used Open-source Project for a Network Data Access Protocol (OPeNDAP) software;
- robust multifile movement between various storage systems, via the development of DataMover technology;
- standardization of data formats;
- XML-based metadata schemas and tools for discovery, use, analysis, managing, and sharing of scientific data;
- security and access controls, supporting both low-cost user registration and authentication for purposes of auditing and highly secure access control for more sensitive operations;
- development of a Web-based portal enabling interactive searching of ESG data holdings, browsing of associated metadata, and retrieval of datasets that may be large.

Our first official release of ESG to the greater climate community began in Spring 2004. The portal is being used to deliver to the community data produced by the Intergovernmental Panel on Climate Change (IPCC) assessment program. With this first production release, ESG's vision of providing the foundation for the next-generation data publication, analysis applications, Web-based portals, and collaborative problem-solving environments for climate research [9] has been demonstrated.

### A. ESG Web Portal

We have developed the ESG Web portal, which provides the entry point to ESG data holdings. Its purpose is to allow climate scientists from around the world to authenticate themselves to the ESG services and to browse, search, subset, and retrieve data. The ESG portal uses Grid-enabled ESG services to associate metadata with datasets, determine data location, and deliver data to the location desired by the user. The portal includes an "Aggregated Data Selection" option that allows the user to select a variable and a subregion and time and level ranges.

### B. Metadata

The role of data descriptions, i.e., metadata, is critical in the discovery, analysis, management, and sharing of scientific data. We have developed standardized metadata descriptions and metadata services to support this important functionality.

ESG has developed metadata schema oriented toward the types of metadata generated by atmospheric general circulation and coupled ocean-atmospheric models. These schema are defined in terms of the Network Common Data Format (NetCDF) markup language (NcML) [10] and provide an XML-based representation of metadata that appears in NetCDF encoded simulation data.

The ESG metadata schema are supported by a range of services that are used to provide access to specific metadata values. The ESG Metadata Catalog is based on the Open Grid Services Architecture Data Access and Integration (OGSA-DAI) service [11] and uses relational technology for browsing and searching ESG metadata holdings. We have also deployed a separate metadata service called the Replica Location Service (RLS) [12], [13] to provide information about the location of datasets, allowing data to be replicated and migrated without having to modify entries in the main catalog of NcML data.

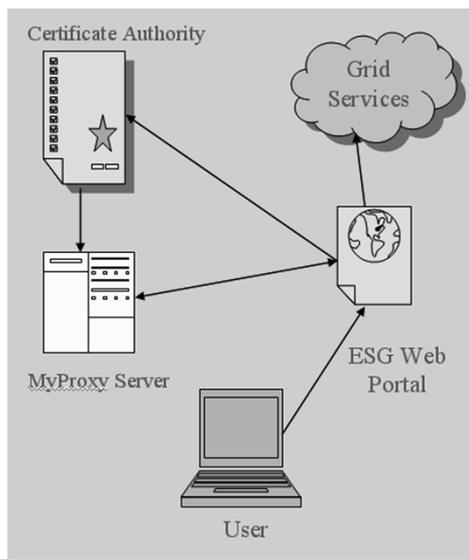

**Fig. 2.** Registration system architecture.

We provide the ability to bridge ESG metadata into dataset catalogs conforming to the Thematic Realtime Environmental Data Distributed Services (THREDDS) specification [14]. THREDDS catalogs are automatically generated from the data description information contained in the OGSA-DAI database and the location information retrieved from the distributed RLS databases.

*C. Security*

ESG's large and diverse user community and the multiple laboratories that participate in ESG make security a challenging problem. We have devoted considerable effort to analyzing and documenting security requirements. ESG has adopted a security approach that includes user authentication, user registration with the ESG portal, and group-based access control.

The Grid Security Infrastructure (GSI) [15], [16] is used as a common underlying basis for secure single sign-on and mutual user-resource authentication. This machinery provides for secure access control to all ESG resources using public key infrastructure (PKI) with credentials from the DOE Grids Certificate Authority (CA). The ESG is recognized by the DOE CA as an accredited virtual organization and participates in the Policy Management Authority Board.

Because of the overhead associated with obtaining public key credentials, an online registration process is used to access "moderate-quality" PKI credentials without the overhead normally associated with certificate generation. This mechanism allows for a broad base of users to access "public" data while providing an auditing of access by remote users with whom ESG has no existing trust relationship, as required by research sponsors.

ESG has developed a registration system that allows users to register easily with the ESG portal. The registration system architecture is shown in Fig. 2. The focus of this system is ease of use. We developed portal extensions (CGI scripts) that automate user registration requests. The system solicits basic data from the user, generates a certificate request for the ESG CA, generates a certificate and stores it in the MyProxy service, and gives the user an ID/password for MyProxy. The registration system also has an administrator's interface that allows the CA administrator to accept or reject a user request. The most important benefits of this system are that users never have to deal with certificates and that the portal can get a user certificate from the MyProxy service when needed.

Shared data access requires the ability to specify and enforce group-based access control for individual datasets. We have implemented a prototype authorization service by integrating the Community Authorization Service (CAS) [17] into the GridFTP [18] server used to implement OPeNDAP-g [19] data queries. This system supports group-based access control based on community specified policy stored in the CAS server's policy database.

*D. Data Transport and Access Services*

OPeNDAP is a protocol and suite of technologies that provide flexible, easy access to remote data. Based on the widely accepted community OPeNDAP software, ESG has extended OPeNDAP to create OPeNDAP-g, a Grid-enabled OPeNDAP server and clients that support GSI security [15] and can use GridFTP [20] as a data transport mechanism.

In addition, seamless joining of data stored in different OPeNDAP servers ("aggregation") is now part of the ESG production release. ESG has also built a prototype end-to-end data access and transport system demonstrating the use of CDAT [3] and NCL [6] to access remote datasets using the GridFTP protocol via the OPeNDAP-g client. Continued work will be necessary before this goes into production, including support of aggregation, but the goal is that this OPeNDAP-g client software will be the foundation for distributed climate applications to gain access to ESG resources.

*E. Multiple File Transport*

ESG has developed tools for the efficient and reliable transfer of multiple files among storage sites. The transport of thousands of files is a tedious, error-prone, but extremely important task in climate modeling and analysis applications. The scientists who run models on powerful supercomputers often need the massive volume of data generated to be moved reliably to another site. Often the source and destination storage systems are specialized mass storage systems, such as the High Performance Storage System (HPSS) at the National Energy Research Scientific Computing Center (NERSC), Berkeley, CA, or the Mass Storage Systems (MSS) at NCAR. During the analysis phase, subsets of the data need to be accessed from various storage systems and moved reliably to a destination site where the analysis takes place. The automation of the file replication task requires automatic space acquisition and reuse; monitoring the progress of thousands of files being staged from the source mass storage system; transferring the files over the network; and archiving them at the target mass storage system. We have leveraged the software developed by the Storage Resource Manager (SRM) [21] project to achieve robust multifile transport for ESG. SRMs are software components that are placed in front of storage systems, accept multifile requests, manage the incremental access of files from the storage system, and then use GridFTP to transport the files to their destination.

Two important results related to multiple file transport were achieved in the first phase of the ESG project. The first was motivated by a practical need to Grid-enable the MSS at NCAR; that is, to be able to store files into and get files from the MSS directly from remote sites. To achieve this goal, we adapted a version of an SRM to NCAR's MSS. This allowed multifile movement between various sites, such as LBNL/NERSC and NCAR. This system has been used repeatedly to move reliably thousands of file between ESG sites. Second, reliability is achieved in the SRM by monitoring the staging, transfer, and archiving of files and by recovering from transient failures. A Web-based tool was deployed to dynamically monitor the progress of the multifile replication.

In response to ESG user requirements for the ability to move entire directories in a single command, a software module called the DataMover [6] was developed. The DataMover interacts with the source and destination storage systems through SRMs to setup the target directory and instructs the SRMs to perform the robust multifile movement.

The DataMover and an SRM are used by the ESG portal to copy user-requested files into the portal's disk space from the files' source locations, which may be remote storage systems. The user is then able to access those files directly from the portal. This capability makes it possible for the portal to act on behalf of a user to get files from remote sites without requiring the user to have direct access to those sites.

### F. Monitoring

We have built monitoring infrastructure that allows us to keep track of the state of resources distributed across seven ESG institutions. Monitoring is required to provide users with a robust and reliable environment they can depend upon as part of their daily research activities. We monitor the hardware and software components that make up the ESG and provide status information to ESG operators and users. Our monitoring infrastructure builds upon the Globus Toolkit's Grid information services [22].

### G. Ontologies for ESG Metadata

We developed a prototype ESG ontology [23], [24] that is available on the ESG portal. The ontology contains the following classes:

- scientific investigation (with subclasses simulation, observation, experiment, analysis);
- datasets (with subclasses campaign, ensembles);
- pedigree (with subclasses identity, provenance, project);
- service;
- access;
- other.

Relationships supported by the ontology include isPartOf, isGeneratedBy, isDerivedFrom, hasParameter, and usesService.

While climate simulation data are the current focus of ESG, the *scientific investigation* metadata defined by our ontology accommodates other types of scientific investigation, for example, observational data collected by oceanographers. *Dataset* metadata includes time and space coverage and other parameters. *Pedigree* or *provenance* metadata trace the origins of a dataset and the successive operations that were performed by recording the conditions under which a dataset of interest was produced; for instance, provenance describes what models, what versions of a model, and what software and hardware configurations were used for a run of interest. *Service* metadata associates earth science data formats with servers providing functionality such as subsetting in coordinate space, visualization, and expression evaluation.

The ESG ontology clearly separates metadata potentially reusable by other scientific applications (such as project information) from metadata specific to an application. The iterative work of detailed concept definition and the rigor needed for specifying relationships between entities required in ontology authoring substantially improved the ESG schema.

### H. Current ESG Data Holdings

One of the key goals of ESG is to serve a critical mass of data of interest to the climate community. We are steadily increasing the amount of data provided to the climate community through ESG.

In the first phase or the project, ESG data holdings consisted of a small set of CCSM [2], [25] and Parallel Climate Model (PCM) [26] data. The number of early users accessing these data via our prototype ESG system was deliberately kept small and ranged from one to five concurrent users.

By the end of 2003, PCM data totaled approximately 100 TB at several sites, including NCAR, the Program for Climate Model Diagnosis and Intercomparison (PCMDI), NERSC, ORNL, and LANL. About 65 users requested accounts at PCMDI to access the PCM data holdings there, representing around 15 science projects, including university researchers, DOE collaborators, and various groups interested in climate simulation on a regional scale.

In 2004, NCAR focused on publishing CCSM data, while LLNL publishED climate simulation data generated by the major climate modeling groups worldwide for the Intergovernmental Panel on Climate Change (IPCC) Fourth Assessment Report (FAR). The IPCC, which was jointly established by the World Meteorological Organization (WMO) and the United Nations Environment Program, carries out periodic assessments of the science of climate change, which serve governments and others interested in the potential impact of future climate change. Fundamental to this effort is the production, collection, and analysis of climate model simulations carried out by major international research centers. Although these centers individually often produce useful analyses of the individual results, the collective model output from several modeling centers often yields a better understanding of the uncertainties associated with these models. Toward this end, the WMO's Working Group on Coupled Modeling (WGCM), in consultation with the IPCC leadership, coordinates a set of standard climate-change simulations carried out by the major modeling centers. In order for this tremendous commitment of computational and scientific effort to be of maximum benefit, it is essential that a wide community of climate scientists critically analyze the model output. The IPCC and WGCM have, therefore, asked PCMDI to collect model output data from these IPCC simulations and to distribute these to the community via ESG.

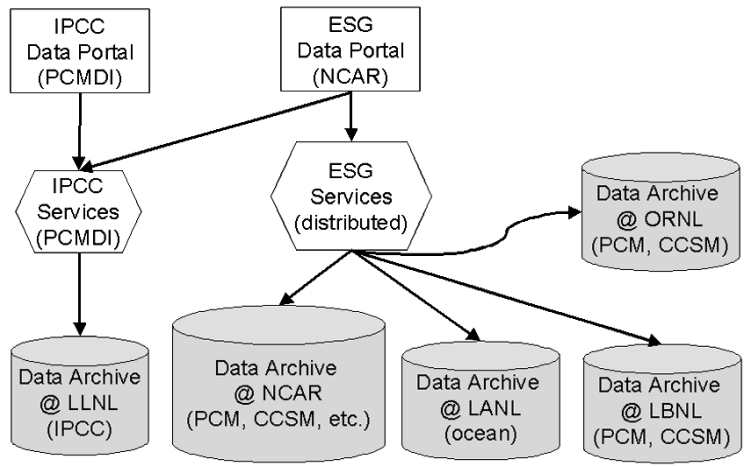

Fig. 3. Schematic of ESG data portals, services, and archives.

By late 2004, the CCSM and IPCC model runs totaled approximately 20 TB of data. CCSM is one of the major climate simulation efforts worldwide, so interest in this data from the scientific community and others will be significant, on the order of hundreds of scientists, researchers, policy makers and others. ESG is providing access to this considerable store of data.

*I. Deployment of Portals, Services, and Archives*

The critical role of the IPCC data and the commitments made by LLNL's PCMDI to provide community access to that data led us to adopt the overall deployment strategy depicted in Fig. 3.

At the lowest level, data are maintained in *archives* at PCMDI (for IPCC data) and NCAR (for other climate simulation data, e.g., from CCSM and PCM). Above these, *services* enable remote access to data archives by maintaining metadata used for data discovery, implementing authentication and authorization, and so on. A copy of these services, located primarily at NCAR, is available for ESG-wide use, and another copy at PCMDI will provide access only to IPCC data. (Currently, the software has been installed at PCMDI, but the site is not yet supporting users of IPCC data.) Similarly, there are two Web *data portals* for data access: one based at PCMDI for IPCC data only, and an ESG-wide portal that will provide access to IPCC, CCSM, PCM, LANL, and eventually other data. The NCAR and PCMDI instantiations of the data portal and services are based on the same code, configured differently to address ESG and IPCC requirements.

*J. External Collaborations*

Where appropriate, we have developed ESG tools and services in collaboration with national and international groups. Currently, ESG is in close collaboration with the British Atmospheric Data Centre in the U.K., the National Oceanic and Atmospheric Administration (NOAA) Operational Model Archive and Distribution System (NOMADS), the Earth Science Portal (ESP) Consortium, THREDDS [14], NOAA's Geophysical Fluid Dynamics Laboratory, and the OPeNDAP project. We have also held discussions with the Linked Environments for Atmospheric Discovery (LEAD) project [28], the Geosciences Cyberinfrastructure Network (GEON) project [29], and the Earth System Modeling Framework (ESMF) [27].

V. FUTURE WORK: INCREASING THE UTILITY OF ESG TO THE CLIMATE MODELING COMMUNITY

In this section, we discuss future work in the ESG project. Over the next two years, the goal of the ESG collaborative is to increase the amount of data made available to scientists through the ESG portal and to enhance the functionality, performance, and robustness of ESG components.

*A. Additional Deployment of Data Archives, Services, and Portals*

The ESG data services and a Web data portal have been installed at PCMDI to support IPCC data distribution there, and they will soon be registering users who want to access the data. Eventually, we plan to include more sites as data archives providing climate datasets to ESG users, including LBNL and ORNL.

*B. Web Portal and Overall System Integration*

The ESG data portal integrates the different ESG services (authentication and authorization, metadata, data processing, and data transport). We will revise the existing Web portal as required and support new user communities (e.g., ocean modelers) as needed, either by providing restricted-access areas on the same portal or by developing customized virtual portals for each community. We will also be engaged in formal usability testing with select members of the research community.

### C. Security Services

The current ESG security architecture deals primarily with authentication of users to the various Grid resources (Grid services, storage systems, etc.). Several key points remain to be addressed, including authorization (or role-based access control to the resources) and support of specific site requirements, such as one-time passwords (OTPs).

*1) Authorization:* Our authorization infrastructure will build upon the Community Authorization Service (CAS) [17] being developed as part of the Globus Toolkit. Building on the Globus Toolkit Grid Security Infrastructure (GSI) [16], [30], CAS allows resource providers to specify course-grained access control policies and delegate fine-grained access control policy management to the community, which specifies its policies via the CAS. Resource providers maintain ultimate authority over their resources but are spared day-to-day policy administration tasks (e.g., adding and deleting users, modifying user privileges).

ESG distinguishes between two types of authorization: file-oriented and services-oriented. *File-oriented authorization* makes access control decisions based on file access permissions for groups of users. The same data on the same server could be accessed by one group of users but inaccessible for another group. *Service-oriented authorization* makes access control decisions based on a group's permission to access particular services. For example, some privileged users might be allowed to access data efficiently from hierarchical storage systems using the DataMover service, while less powerful users would be able to access that data only through the OpenDAP-g server.

For both types of authorization, users will be given their access rights at their initial sign-up with the ESG portal via the ESG registration system. Any further changes to user access rights would be done by the resource providers using special tools (to be developed).

*2) One-Time Password (OTP) Authentication:* There are various options for supporting the use of one-time passwords, including an online CA that generates GSI credentials automatically for a user who has authenticated with the OTP system (thus allowing them to access other ESG sites without further authentication). Another possibility is for GSI credentials generated by an online CA at one site following OTP authentication to be accepted at another site with the same requirement. A third option would include changing GSI credentials to specify whether authentication to generate the user certificate used OTP. We will investigate these options and implement the one that best fits ESG.

### D. Metadata Schema and Services

ESG has focused strongly on metadata. Remaining tasks include the following.

*1) Metadata Schema:* From the start, it has been our aim in ESG to develop a system that has the potential for longevity and interoperation with other emerging data systems worldwide. New standards, like ISO-911, are emerging that can facilitate interoperation of multiple data systems, and we plan to spend some effort evolving our metadata in this direction. This work will be undertaken primarily by LLNL and NCAR, but in concert with a number of other projects (e.g., British Atmospheric Data Center [31], THREDDS [14], etc.) that face the same future needs. The metadata changes required by ESG researchers will be incorporated into the existing OGSA DAI-based metadata service.

*2) Metadata Catalog Support for Virtual Data:* Work is required to extend our metadata catalogs to support virtual data definitions, so that data producers can define virtual datasets and data consumers can discover and request virtual datasets. We intend to work with NcML as our virtual data definition language.

*3) Multiple Metadata Catalog Services:* We will implement a replicated metadata catalog to avoid a single point of failure. The initial goal of this work will be to improve reliability, but we also note that replication can improve performance by load balancing of the metadata query workload. Initially, we plan to use the OGSA-DAI metadata catalog at NCAR as the master catalog and do periodic updates to a second catalog located at Lawrence Livermore National Laboratory. We will deploy more general distributed metadata services as they are developed. We have also done some exploratory work in using the Open Archive Initiative (OAI) protocols [32] to accomplish this function, and we will evaluate the relative effectiveness and level of effort required to do this.

*4) Browse, Search, and Query:* We will support richer metadata catalog query capabilities for ESG scientists.

*5) Federation:* We will provide for the interoperation of two heterogeneous metadata services: our ESG Metadata Services and the THREDDS catalogs. We will provide simple distributed queries across the two catalog types.

### E. DataMover Services

Our work thus far with the DataMover application [6] has resulted in a unique capability that copies directories (effectively providing the equivalent of a Unix "rcp—r" copy command) across a heterogeneous online/archival storage environment, including NCAR's locally developed MSS. Strong security—a critical prerequisite for such actions—is required and, thus, the use of DataMover is largely restricted to the "power users" (i.e., users that have formal accounts at all sites involved in data transfer). But this leaves out the average user who may want to use the Web portal to locate and identify a large number of files that must be moved back to a local system. The Web portal provides a workable interface for the selection of a small number of files (limited to a few gigabytes). However, if a user wants a large number of files, then we need a different sort of capability, because such transfers require automation to deal with queuing,

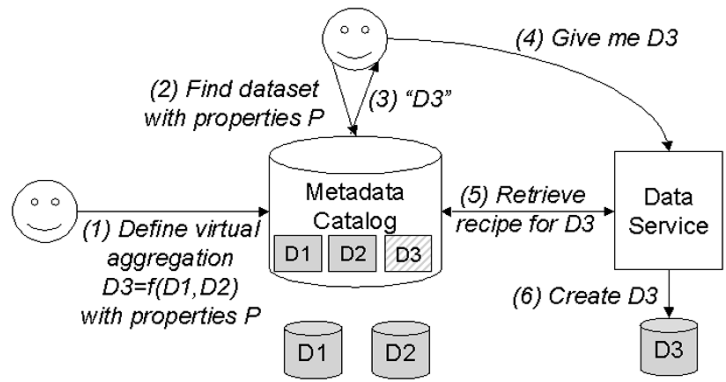

Fig. 4. Steps involved in defining, discovering, requesting, and instantiating a virtual dataset.

management of cache disk space, network problems, and recovery from transient failures.

We will continue to refine, enhance, and increase the robustness and scaling of the DataMover tool. Beyond that, we will evolve the current DataMover into a "DataMover-Lite" application that will work seamlessly with the Web portal, allowing a user to make a selection of a large number of files and then trigger the launch of this application. We plan to provide data movement capabilities for two types of users which we refer to as "casual users" and "frequent users." Casual users are those that do not have Grid credentials, and frequent users are those that are willing to go through the process of acquiring Grid credentials to achieve more efficient transport of files. For casual users, the requested files will have to be moved to the disk cache managed by the Web portal first, and only then can they be "pulled" by the users to their systems. For frequent users, we plan to develop a client-side "DataMover-Lite" that will "pull" the files directly from their source location to the user's location, avoiding transport through the Web portal's disk cache. This more efficient method of file transport is necessary when moving a large volume of data. Also, to conform to security policies, "DataMover-Lite" will work from behind the firewall and adapt to the local security policy.

*F. Aggregation Services and OPeNDAP-g*

Additional development of OPeNDAP-g will enhance virtual data services for the Web portal, native client access, performance, and robustness.

*1) Clients:* ESG clients include the Web portal and user desktop applications such as CDAT [7] and NCL [8]. We plan to complete the netCDF client interface to these applications, providing full functionality of the API and transparently handling aggregation and subsetting (i.e., Virtual Data Services) to return useful data to users. It will be necessary for client applications to operate with various implementations of ESG security and to utilize an OPeNDAP-constrained URL to access data. In some cases, the client may need to generate these URLs, thus requiring access to the ESG catalogs.

*2) Servers:* We will migrate the OPeNDAP-g services to the latest release of the Globus Toolkit Web services-based components and the new striped GridFTP server. Integration of ESG security models and connections to catalog services from OPeNDAP-g client interface are expected to require modifications to the OPeNDAP-g core libraries. In addition, we plan to complete the integration of the data access and transport with RLSs and SRMs for hierarchical storage systems.

*G. Monitoring*

We will provide enhanced monitoring capabilities to improve the robustness of the ESG infrastructure. These will include monitoring a larger number of ESG services and components and providing better interfaces for querying historical information about resource availability.

*H. Virtual Data Support*

We will provide support for virtual data in the next phase of ESG. Fig. 4 depicts the tasks involved in publishing, discovering, and accessing a virtual dataset, which include the following.

1) A data provider publishes to an ESG metadata catalog a definition of a new dataset, indicating its name ("D3"), associated metadata ("P"), and either its constituent component files (if a physical dataset) or its definition (if a virtual dataset, as here: "f(D1, D2)").
2) A client (user or program) issues a query to the metadata catalog for datasets of interest.
3) The names of any datasets matching the user query are returned. In the figure, this query happens to match the properties associated with the virtual dataset D3, and so the name D3 is returned.
4) A client requests that dataset by requesting the "Virtual Data Service" (shown as "Data Service" in the figure) for the dataset D3 or any subset of D3.
5) The Virtual Data Service retrieves the recipe for D3 from the metadata catalog.
6) The Virtual Data Service instantiates D3 (or a specified subset of D3) by fetching appropriate data from D1 and D2 and assembling those pieces to create a new dataset D3.

Finally, the Virtual Data Service returns the requested data to the user (step not shown).

The Virtual Data Service may also publish the location of the new physical dataset into the metadata catalog so as to accelerate the processing of subsequent requests for the dataset (step also not shown).

## VI. RELATED WORK

ESG has worked closely with several existing efforts. These include the DOE Science Grid Project [33], whose work on authentication infrastructure and related security issues has been particularly useful to ESG. ESG uses the DOE Science Grid Certification Authority as the basis for its authentication.

ESG has also worked closely with Unidata [34] on several initiatives, including joint development of the NcML specification [10] that provides a standard XML encoding for the content of netCDF files. NcML has been extended to support coordinate system information, aggregation/subsetting logic, and GIS interoperability. NcML is still being developed and is being established as a community standard. ESG also makes use of the THREDDS specification developed within Unidata.

The ESG collaborative includes members from the Globus Alliance [35]. ESG has made extensive use of Globus Toolkit components. For example, ESG acted as an early adopter and stringent beta tester of the GridFTP code [18], identifying a number of subtle errors in the implementation. ESG has collaborated with Globus on compelling technology demonstrations involving data movement at close to 1 GB/s over wide area networks. ESG also made use of the GSI [16], the RLS [12], [13], remote job submission capabilities, monitoring and information systems, and the CAS [17] provided by the Globus Toolkit.

ESG also includes members of the SRM Project at LBNL [21], [36]. SRMs provide dynamic storage management and support for multifile requests. The LBNL team adapted the SRM developed for HPSS to the MSS at NCAR. As part of ESG, the SRM team also developed DataMover, providing the ability to move entire directories robustly (with recovery from failures) between diverse mass storage systems at NCAR, LBNL, and ORNL, as well as disk systems at NCAR and LLNL. The DataMover has been used repeatedly by members of the ESG team to move thousands of files robustly between mass storage systems at these sites.

The ESMF [27] is a NASA-funded project that has engaged staff at NCAR, NASA, and DOE laboratories to develop a standard framework for the construction of modular climate models. In a next phase, the scope of ESMF and ESG both expand to embrace problem-solving environments for earth system researchers. We see strong opportunities for mutually beneficial interaction between ESG and ESMF as models are extended both to access remote data and to publish data.

A number of scientific projects face challenges similar to those being explored in ESG. The Grid Physics Network (GriPhyN) project [37] uses grid technologies to support physicists, with emphasis on support for virtual data, data management and workflow management. The Particle Physics Data Grid [38] also employs grid technologies to support physics research. The LEAD project [28] is developing scientific and Grid infrastructure to support mesoscale meteorological research. The GEON project [29] supports geoscientists with an emphasis on data modeling, indexing, semantic mediation, and visualization.

ESG technology is based on Globus Toolkit and SRM middleware. Other scientific grid projects use the Storage Resource Broker (SRB) middleware [39], [40]. Unlike the layered approach taken by the Globus Toolkit, SRB provides tightly integrated functionality that includes extensive data management capabilities, including support for metadata, organizing data in collections and containers, and maintaining consistency among replicas.

## VII. CONCLUSION

The increasingly complex datasets being produced by global climate simulations are fast becoming too massive for current storage, manipulation, archiving, navigation, and retrieval capabilities. The goal of the ESG is to provide data management and manipulation infrastructure in a virtual collaborative environment that overcomes these challenges by linking distributed centers, users, models, and data. An important role of the ESG project is to provide a critical mass of data of interest to the climate community, including CCSM, PCM, and IPCC simulation model output.

Over the past three years, the ESG project has made considerable progress toward the goal of a community ESG. We have developed a suite of technologies including standard metadata schema; tools for metadata extraction; metadata services for data discovery; security technologies that provide authentication, registration and some authorization capabilities; a grid-enabled version of "OPeNDAP-g" for high-performance data access; robust multiple file transport using SRM and DataMover; a monitoring infrastructure; and the development of a Web portal for interactive user access to climate data holdings. To date, we have catalogued close to 100 TB of climate data, all with rich scientific metadata.

In the next phase of ESG, we will increase the utility of ESG to the climate modeling community by expanding both the data holdings that we provide and the capabilities of the system. Over the next two years, the ESG will provide enhanced performance and reliability as well as richer authorization, metadata, data transport, and aggregation capabilities and support for virtual data.

Authors' photographs and biographies not available at the time of publication.